# Mitigating Data Centers Load Risks and Enabling Grid Support Functions through Grid-Forming Control

Yousef Abudyak, Mohsen Alizadeh, and Wei Sun
Department of Electrical and Computer Engineering
University of Central Florida (UCF)
Orlando, USA
yousefhusamyousef.abudyak@ucf.edu, mohsen.alizadeh@ucf.edu, sun@ucf.edu

*Abstract*— The rapid growth of hyperscale data centers driven by Large Language Models and Artificial Intelligence workloads has introduced new challenges for power systems. These facilities experience abrupt power variations during model training and check-point-saving events, causing voltage deviations and frequency disturbances. Moreover, they operate as passive loads that draw power without offering any grid support. This paper presents an integrated architecture that combines Battery Energy Storage Systems (BESSs) within data centers using Grid-Forming inverters to provide active grid-support functions. Simulation results through MATLAB/Simulink demonstrate accurate power reference tracking under dynamic loading, with eight coordinated BESS units supplying instantaneous power during training and saving conditions. Under single-phase voltage depression near the data center bus, the BESS delivered reactive power support similar to a Static Synchronous Compensator. During grid disconnection, seamless islanded operation was achieved with stable voltage, frequency, and continuous power delivery at the data center bus.

## I. Introduction

The Bulk Power System (BPS) is experiencing a profound change on both the generation and load sides, reshaping its operational characteristics and dynamic behavior. On the generation side, the accelerated deployment of renewable energy resources and inverter-based technologies is primarily driven by global decarbonization goals and economic incentives associated with zero-emission initiatives [1]. Conversely, on the load side, the proliferation of large, power-intensive facilities, most notably data centers supporting Artificial Intelligence (AI) training and Large Language Model (LLM) operations, has introduced unprecedented demand dynamics [2].

Several technical reports have recently examined the role of hyperscale data centers in BPS operations [2-6]. North American Electric Reliability Corporation (NERC) identified data centers and other power electronic loads as drivers of voltage-sensitive events, emphasizing their ramp rates, unpredictability, and fast interconnection timelines as emerging reliability risks [6]. The Western Electricity Coordinating Council echoed these findings, warning that rapid data center-driven load growth mirrors the earlier inverter-integration crisis and requires immediate regulatory and technical actions [2]. Elevate and Grid Lab Guidance highlighted the need for transparent interconnection processes, improved modeling, and coordination between transmission and distribution planners [5]. In contrast, [3] demonstrated that controlled demand response from data centers can significantly reduce capacity needs, while the Idaho National Laboratory clean energy case study showed that firm nuclear generation remains the most dependable and cost-effective option for powering large-scale facilities compared to variable renewables [4]. Together, these reports reveal both the operational risks and the potential system value of large data centers when properly integrated into grid planning and reliability frameworks.

Figure 1 illustrates the measured active-power profile of a 50-MW data center facility, revealing distinct ramp-up and ramp-down transitions linked to AI workload cycles. The rapid ramp-up corresponds to the initiation of deep-learning training phases, when thousands of Graphics Processing Units (GPUs) and accelerators operate in parallel, drastically increasing electrical demand. The subsequent ramp-down occurs during checkpoint saving, when computations pause and GPU utilization drops sharply. This critical aspect, recently introduced and emphasized by both GridLab and NERC, underscores how AI-driven data center loads exhibit unconventional, burst-type behavior that can excite grid oscillations and challenge system stability if not properly mitigated [5, 6]. This observed load pattern will be used as the representative input in the simulation framework to evaluate how the proposed topology mitigates rapid ramping effects and stabilizes the system response. By replicating these training–checkpoint cycles, the study validates the effectiveness of the control scheme under realistic, AI-driven load variability conditions.

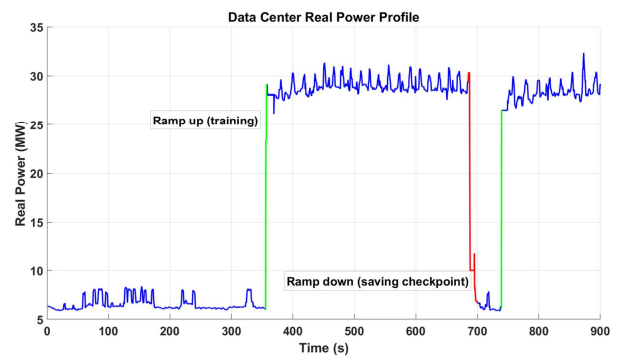

Figure 1. Real power profile of the data center facility, illustrating steady-state fluctuations (blue), ramp-up periods during training (green), and ramp-down periods for saving checkpoints (red).

Based on available data [7, 8], a comparable growth trend can be observed between the expansion of renewable energy capacity, primarily composed of Inverter-Based Resources (IBRs), and the increasing power demand from data centers. Figure 2 illustrates this relationship over the period 2018–2024, showing the growth of renewable energy capacity in the United States (including battery storage systems and wind generation units rated above 1 MW) alongside the total data center electricity demand across all facility types.

The concurrent growth of IBRs and data centers highlights the need for their coordinated integration within the BPS. Rather than operating independently, data centers can be directly supplied by storage-based IBRs utilizing advanced Grid-Forming (GFM) control architectures [9]. This coordinated configuration enables renewable resources to dynamically support the rapidly growing and variable data center demand while mitigating the operational and stability challenges that arise when these facilities are separately connected to the grid. Also, through such integration, data centers can replace conventional diesel generators with sustainable backup power sources based on storage-integrated IBRs [10].

In addition to improving on-site reliability, the same configuration allows IBRs to provide essential grid-support functions, including reactive power support, voltage and frequency regulation, and islanding mode operation [11]. Moreover, replacing grid-following inverters with GFMs for the associated IBRs further enhances the system's ability to maintain voltage and frequency stability under weak-grid conditions [12]. In this manner, each data center transitions from being a potential burden on the BPS to an active asset that enhances overall system stability and resilience.

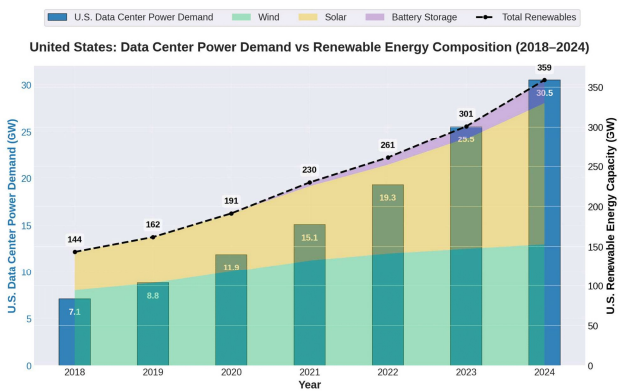

Figure 2. U.S. data centers power demand and renewable energy capacity growth (2018–2024).

The rest of this paper is organized as follows. Section II discusses the structure of the conventional and proposed data centers. Section III presents the control strategies employed with energy storage–based IBRs to enable grid-supportive functions. Finally, Section IV provides the simulation results, discussion, and key observations, followed by the concluding remarks.

## II. CONVENTIONAL AND PROPOSED DATA CENTER ARCHITECTURES

In this section, the conventional and proposed architecture of data centers are discussed. Although various types of data centers exist, the focus here is on hyperscale facilities, given their significant power demand and increasing impact on the BPS [6].

### A. Conventional Data Center Architecture

A hyperscale data center employs a highly redundant electrical design to ensure uninterrupted power under any single failure or maintenance event. The facility connects to the high-voltage grid through N + 1 transformers, stepping power down to Medium Voltage (MV) and distributing it to multiple buildings divided into halls and pods. Each pod includes its own transformer, diesel generator, and dual uninterruptible power supply (UPS) systems that provide instant ride-through power during faults until the generator starts and stabilizes [13]. Depending on uptime requirements, data centers typically adopt either N+1 or 2N redundancy [14]. N+1 designs provide one independent backup for each set of critical components such as transformers and generators, while 2N designs fully duplicate all power paths for maximum resilience. Rated 3 (Tier III) facilities are concurrently maintainable, generally using N+1 for supply and generation and 2N for UPS and distribution systems. This configuration eliminates single points of failure and ensures continuous operation consistent with Tier III reliability standards [15]. Figure 3 shows the electrical system diagram of a hyperscale data center with two buildings (adapted from Microsoft building site).

### B. Proposed Data Center Architecture

As discussed previously, the proposed solution aims to supply the facility with the required power without fully depending on the grid. However, the facility remains interconnected with the grid to enable grid-supportive functions, allowing it to assist during events that demand fast frequency response or Static Synchronous Compensator (STATCOM)-type reactive power support. This configuration also enhances system resilience by enabling the facility to participate in restoration and black-start events when required, however this is out of the scope of this paper [16].

The criteria for connecting the Battery Energy Storage Systems (BESSs)-based GFM inverter in this study are derived from the design framework presented in [15], which applies gas engine systems to hyperscale data centers. This reference design was chosen for its comparable power rating and structural similarity to the interconnection configuration described in Section II.A. Figure 4 shows the connection of the BESS to the data center facility.

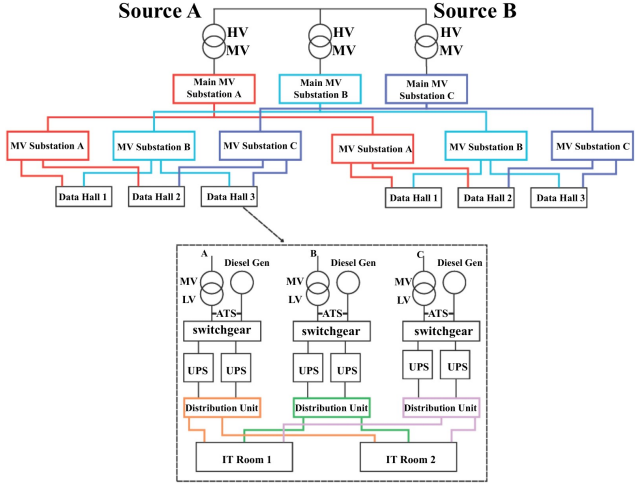

Figure 3. Electrical system diagram of a hyperscale data center.

It is worth mentioning that the type of batteries used in such facilities is critical, such as lithium-ion systems, while dominant today, pose significant challenges for large-scale, long-duration applications [17]. Their fire risk, limited storage duration, and dependence on a heavily constrained, China-centered supply chain have driven interest toward long-duration energy storage alternatives. Emerging technologies such as thermal and flow batteries offer safer and more sustainable options for data center integration [18].

### III. BESS–GFM ARCHITECTURE AND CONTROL DESIGN

The BESS-based GFM inverter utilizes a hierarchical control architecture to enable voltage formation, power regulation, and seamless grid interaction at the Point of Common Coupling (PCC). As shown in Fig. 5, the converter is connected through an L – C filter composed of inverter-side inductance $L_f$, shunt capacitance $C_f$, and equivalent grid-side inductance $L_g$. The base impedance is

$$Z_{Base} = \frac{V_{n,SB}^2}{S_{n,SB}}, \quad (1)$$

where $V_{n,SB}$ and $S_{n,SB}$ denote the nominal voltage and apparent power of the STATCOM–BESS, respectively. The filter components are determined as [16]

$$L_f = \frac{0.15\, Z_{Base}}{2\pi f}, \quad C_f = \frac{0.15}{2\pi f\, Z_{Base}} \quad (2)$$

The capacitor $C_f$ provides reactive-power support and filters harmonics, while $L_f$ limits high-frequency current components and improves coupling with the grid.

Measured voltages and currents $(v_{abc}, i_{abc})$ are converted to the $dq$-reference frame and processed through cascaded voltage and current control loops. The voltage PI controller $(k_{pv} + k_{iv}/s)$ generates the reference currents $(i_d^*, i_q^*)$, while the current controller $(k_{pi} + k_{ii}/s)$ tracks them to produce voltage references $(v_{rd}^*, v_{rq}^*)$ for the PWM stage. Feed-forward compensation $F$ and the cross-coupling terms $(\pm \omega L_f, \pm \omega C_f)$ enhance dynamic response and reduce output impedance. Setting $V_{cq}^* = 0$ decouples the control channels such that $i_d$ governs active power and $i_q$ governs reactive power [19].

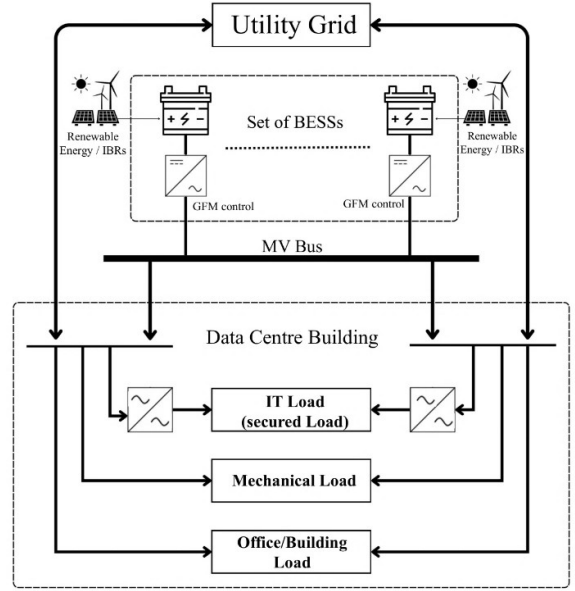

Figure 4. Simplified schematic of the proposed GFM–based data center power architecture, showing the bidirectional interaction between the utility grid, BESSs under GFM control.

The outer power-control layer calculates the instantaneous active and reactive powers as

$$P_{meas} = v_{d,meas} i_{d,meas} + v_{q,meas} i_{q,meas} \approx v_{d,meas} i_{d,meas}, \quad (3)$$

$$Q_{meas} = -v_{d,meas} i_{q,meas} + v_{q,meas} i_{d,meas} \approx -v_{d,meas} i_{q,meas}$$

The measured quantities are filtered through low-pass filters with cutoff frequencies $\omega_p$ and $\omega_q$. The filtered values are then applied to the droop relations defined by $K_p$ (frequency–power) and $K_q$ (voltage–reactive-power)

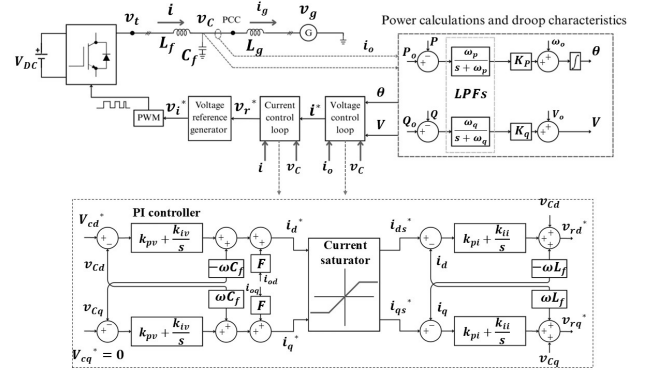

Figure 5. Overall control architecture of the GFM inverter with voltage–current double loop and droop characteristics.

coefficients to generate the reference angle function $\theta$ and voltage magnitude $V$.

This hierarchical configuration allows autonomous frequency and voltage regulation, seamless transition between grid-connected and islanded modes, and fast inertial and reactive-power support. Based on the PI controller tuning and the appropriate choice of physical filter components and BESS rating, the converter can additionally provide advanced grid-support functions such as STATCOM-like reactive compensation, grid restoration participation, and even black-start capability. Consequently, the BESS–GFM structure not only stabilizes voltage and frequency but also enhances overall grid resilience against the dynamic and unpredictable behavior of large data-center loads.

## IV. SIMULATIONS RESULTS AND ANALYSIS

Several simulation scenarios were performed to evaluate and demonstrate the effectiveness of the proposed integration. The results and observations from these scenarios are discussed in this section.

### A. Active Power Support for Data Center

The load profile illustrated in Fig. 2 is adopted in the simulation and directly connected to the grid MV bus. At the same bus, eight GFM-BESS units, each rated at 5 MW, are integrated. Collectively, these inverters provide a total capacity of 40 MW and are designed to share the load equally. Figure 6 shows the behavior of the voltage and frequency before and after the connection of these GFM units.

As observed, when data centers are supplied directly from the grid without an on-site BESS, the voltage experiences a significant drop that inversely follows the load profile. In other words, periods of high power demand from the data center lead to increased current draw, resulting in larger voltage drops across the system. Such voltage deviations may necessitate corrective actions, such as on-load tap changer operations, to restore the voltage to acceptable levels. However, when the BESS units supply the load directly on-site, there is no need to draw current from the grid. In this configuration, the data center can both meet its own demand and provide additional support to the grid. Specifically, it can deliver excess power when available or contribute to mitigating frequency drop events, thereby acting as an active grid-supporting asset rather than a passive load and a potential source of disturbance. In this scenario, the GFM control exhibited a fast and precise dynamic response, effectively tracking the power reference commanded by the data center within milliseconds, as illustrated in Fig. 7. The proposed approach is validated in Simulink environment; however, practical factors such as measurement noise and millisecond-scale communication delays may affect transient performance and require further investigation. With each inverter operating on a 5 MW base, this rapid response significantly improved system frequency behavior. The frequency deviations, which previously reached 59.5 Hz during loading and 60.3 Hz during unloading, were mitigated to 59.8 Hz and 60.15 Hz, respectively, when the GFM-BESS system was employed, as shown in Fig. 6.

### B. Reactive Power Support from Data Center

Instead of relying on a dedicated STATCOM to support the grid, these facilities can contribute to fault conditions by providing reactive power at the MVAr level. To validate this capability, a single line-to-ground fault was applied at a bus located approximately two miles from the data center. As shown in Fig. 8, the fault occurred at second 13 and was cleared at second 18. Upon fault initiation and the resulting voltage collapse at the nearby bus, the GFM-BESS system promptly injected around 5 MVAr of reactive power according to the predefined voltage-droop settings. This immediate response effectively mitigated the voltage depression, restoring the voltage to within the acceptable non-emergency range (0.9–1.1 p.u.). Furthermore, the data center's 6 MW load continued to operate without interruption, exhibiting only minor transient spikes that would be further damped if the UPS systems were included in the simulation model.

### C. Data Centers after disconnection from the grid

As mentioned earlier, these facilities can leverage GFM units to deliver a response similar to that of diesel generators but with significantly faster dynamics in the event of a grid disconnection, as illustrated in Fig. 9.

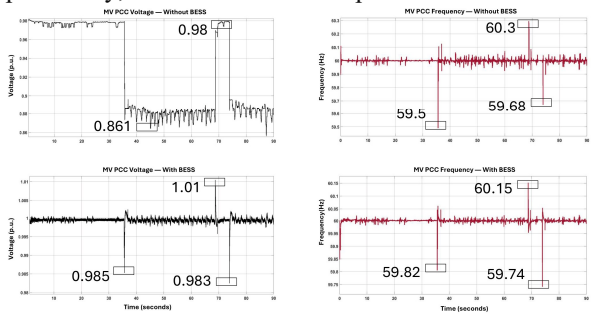

Figure 6.  MV PCC voltage and frequency response at the data center bus. Comparing operation without and with BESS integration.

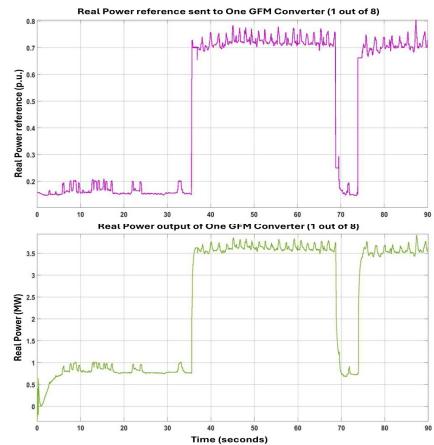

Figure 7.  Real power reference and corresponding real power output when data center is connected with BESS .

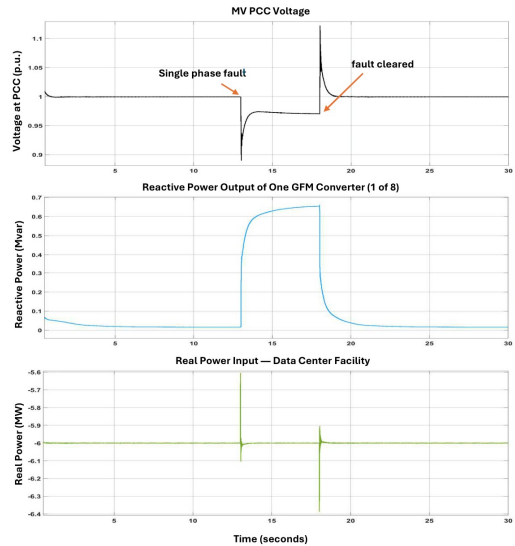

Figure 8.  MV PCC voltage, reactive power output, and real power input to the data center facility during fault conditions.

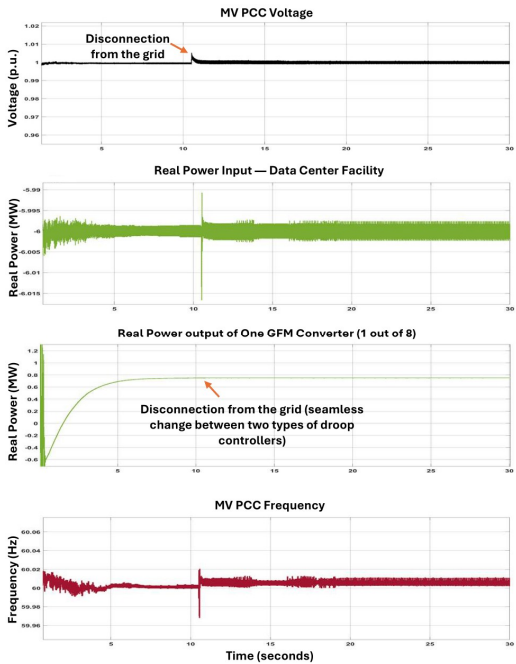

Figure 9.  MV PCC voltage, real power output, real power input to the data center facility, and frequency response before and after islanding.

The results indicate a seamless transition from the grid-tied operation to stand-alone operation, without any violation of voltage or frequency limits. Also, after the disconnection at second 10.5 occurred, and the data center's real power requirements were satisfied with minor spike happened at the disconnection time for microseconds.

## Conclusion

This study demonstrated that integrating storage-based control within data centers can effectively transform them from passive loads into active grid-supporting assets. Simulation results confirmed that the proposed architecture mitigates voltage and frequency deviations caused by AI-driven demand variations, providing rapid active-power response during training and checkpoint cycles. During fault conditions, the system delivered MVAr-level reactive support that maintained the voltage within the acceptable 0.9–1.1 p.u. range, while during grid disconnection, seamless islanded operation was achieved without power interruption. Overall, the proposed integration enhances system reliability, enables fast dynamic response, and strengthens grid resilience. Future work will address coordination with existing protection systems, investigate scalability to multiple data-center facilities, and validation under broader network conditions.

## References

[1] C. Breyer et al., "On the history and future of 100% renewable energy systems research," IEEE Access, vol. 10, pp. 78176–78218, 2022.
[2] An Assessment of Large Load Interconnection Risks in the Western Interconnection, Western Electricity Coordinating Council, Salt Lake City, UT, USA, 2024.
[3] C. Cox, A. Schwartz, and D. Stenclik, Bringing Data Center Flexibility into Resource Adequacy Planning: A Case Study of NV Energy, GridLab and Telos Energy, Washington, DC, USA, 2025.
[4] G. J. Soto et al., Powering Data Centers with Clean Energy: A Techno-Economic Case Study of Nuclear and Renewable Energy Dependability, Idaho National Laboratory, Idaho Falls, ID, USA, 2024.
[5] R. Quint, K. Thomas, J. Zhao, A. Isaacs, and C. Baker, Practical Guidance and Considerations for Large Load Interconnections, Elevate Energy Consulting, Washington, DC, USA, 2025.
[6] Characterization and Risks of Emerging Large Loads, North American Electric Reliability Corporation, Atlanta, GA, USA, Jul. 2025.
[7] Electric Power Annual 2024, U.S. Department of Energy, Washington, DC, USA, 2024.
[8] A. Young, "Data center energy demand to triple by 2030," Avison Young, Chicago, IL, USA, Nov. 13, 2024.
[9] Y. Abudyak, M. H. Rezaei, I. Batarseh, H. S. Rizi, and A. Q. Huang, "Grid-forming control: Utilizing Andronov–Hopf oscillator dynamics with filterless droop characteristics," in Proc. 50th Annu. Conf. IEEE Ind. Electron. Soc. (IECON), 2024, pp. 1–6.
[10] V. Kambhampati, A. van den Dobbelsteen, and J. Schild, "Moving beyond diesel generators: Exploring renewable backup alternatives for data centers," J. Phys.: Conf. Ser., vol. 2929, no. 1, Art. no. 012008, 2024.
[11] A. Alassi, Z. Feng, K. Ahmed, M. Syed, A. Egea-Alvarez, and C. Foote, "Grid-forming VSM control for black-start applications with experimental PHiL validation," Int. J. Electr. Power Energy Syst., vol. 151, Art. no. 109119, 2023.
[12] Y. Li, Y. Gu, and T. C. Green, "Revisiting grid-forming and grid-following inverters: A duality theory," IEEE Trans. Power Syst., vol. 37, no. 6, pp. 4541–4554, Nov. 2022.
[13] D. Patel, J. E. Ontiveros, and D. Nishball, "Data center anatomy: Electrical infrastructure," SemiAnalysis Newsletter, Oct. 14, 2024.
[14] W. Youssef, A. Rajewski, M. Megdiche, and J. Kerttula, Applying Natural Gas Engine Generators to Hyperscale Data Centers, Rueil-Malmaison, France, 2020.
[15] How Are Data Center Tiers Classified and Why Are They Important?, Prasa Infocom, Jun. 29, 2021.
[16] J. K. Schwarzkopf and M. Walti, Control of a Battery Energy Storage and STATCOM for Black-Start Services, 2024.
[17] J. Conzen, S. Lakshmipathy, A. Kapahi, S. Kraft, and M. DiDomizio, "Lithium-ion battery energy storage systems (BESS) hazards," J. Loss Prev. Process Ind., vol. 81, Art. no. 104932, 2023.
[18] Z. Skidmore, "Watt's next? How can batteries be best utilized in the data centre sector?" Data Centre Dynamics, Sep. 3, 2023.
[19] Y. Abudyak, M. H. Rezaei, A. A. B. Abdelnabi, H. S. Rizi, I. Batarseh, and A. Q. Huang, "Grid-forming inverters review: Control, stability, and the next stage with artificial intelligence and digital twins," IEEE Open J. Power Electron., vol. 7, pp. 351–387, 2026.